# The Importance of Aqueous Metabolites in the Martian Subsurface for Understanding Habitability, Organic Chemical Evolution, and Potential Biology

*a white paper for MEPAG SFL-SAG (July 21, 2025)*

Jennifer L. Eigenbrode, NASA Goddard Space Flight Center, jennifer.l.eigenbrode@nasa.gov

Luoth Chou, NASA Goddard Space Flight Center, luoth.chou@nasa.gov

Aqueous metabolites in terrestrial subsurface environments provide critical analog frameworks for assessing the habitability of Martian subsurface ice. On Earth, they play critical roles in sustaining microbial life within soils, permafrost, and groundwater environments and their availability shape microbial community compositions, activity, and adaptability to changes in environmental conditions, enabling communities to persist over millennial timescales. The counterpart to aqueous-soluble organics is the insoluble organic matter pool that makes up the largest portion of organic matter in natural samples and includes most types of organic signatures indicative of biological processes. Employing a range of sample preparation, molecular separation, detection, and imaging techniques enables the characterization of both labile (*i.e.,* soluble and reactive) and recalcitrant (i.e., insoluble, non-reactive; include macromolecules) organic pools. Multiple orthogonal analytical modalities strengthen interpretations of signatures that we associate with biology as we know it and don't know it, by constraining possible abiotic sources, validating measurements across distinct techniques, and ensuring flexibility to interrogate diverse organic chemistries encountered in Martian subsurface environments. This holistic triage approach aligns with the priorities articulated in the Mars Exploration Program Analysis Group's Search for Life - Science Analysis Group (SFL-SAG) Charter for a medium-class Mars mission focused on extant life detection.

**Aqueous metabolites in terrestrial subsurface environments as analogs for Mars.** Leveraging from terrestrial examples, aqueous metabolites, including small, soluble organic compounds that have functional groups that make them more reactive and labile in the environment. Water-soluble metabolites include fatty acids, amino acids, lipids, other organic acids, sugars, and volatile organic compounds that are composed of C, H, and O with lesser N, S, and P, and play critical roles in sustaining microbial life within soils, permafrost, and groundwater environments on Earth. In terrestrial soils, soluble-organic compound availability shapes microbial community composition, activity, and adaptability to changes in environmental conditions, thereby ecosystem productivity and nutrient turnover (Kuzyakov & Blagodatskaya, 2015; Song et al., 2024). In terrestrial permafrost systems, aqueous metabolites accumulate within unfrozen water films and brine inclusions that persist below 0°C due to the colligative effects of dissolved solutes (Amato et al., 2010; Gilichinsky et al., 2007). Compounds such as acetate, formate, and amino acids support basal metabolic activity, maintaining cellular integrity and biochemical function over millennial timescales (Abrov et al., 2021; Stapel et al. 2018; Wilhelm et al., 2012; Murray et al., 2012). Aqueous metabolites largely originate from relic organic matter degradation or low-level microbial metabolic processes. Their presence in permafrost enables microbial adaptation to energy limitation, osmotic stress, and freeze-thaw dynamics within permafrost ecosystems. Similarly, in terrestrial groundwater, aqueous metabolites sustain heterotrophic and chemolithoautotrophic metabolisms in oligotrophic aquifers (Griebler & Lueders, 2009). Low molecular weight small organic acids and amino acids, as well as protein-, lipid-, and lignin-like compounds have been found in deep aquifers and associated with lithoautotrophic ecosystems (Kieft et al, 2018). Aqueous metabolite pools in terrestrial environments are thus fuel for microorganisms (a factor for habitability) and potentially derived from microorganisms (source of life signatures). The presence and compound composition of soluble organic compounds in extraterrestrial subsurface environments is a key measurement for ascertaining the presence and potential for a modern subsurface microbial ecology or past frozen record.

Aqueous metabolites in terrestrial subsurface environments provide critical analog frameworks for assessing the habitability of Martian subsurface ice. On Mars, aqueous metabolites may arise from abiotic processes such as photochemical and radiolytic organic reactions, water–rock interactions, or meteoritic delivery of organics (Fox et al., 2019; Glavin et al., 2020). Hydrated salts and perchlorates detected in the regolith (Hecht et al., 2009) depress the freezing point of water, enabling the formation of thin briny films capable of concentrating and stabilizing metabolites in otherwise frozen environments (Garcia-Descalzo et al., 2020). Such brines, sufficiently sheltered from ionizing radiation at the Martain surface, could maintain physicochemical conditions conducive to metabolic activity, molecular preservation, and potential biosignature retention (Chou, 2019).

**The insoluble organic matter complement.** The counterpart to aqueous-soluble organics is the insoluble organic matter pool, including macromolecular compounds, microbial cells, degraded membranes, and mineral-bound cells, constitutes the dominant organic carbon pool in terrestrial soils, permafrost, and groundwater. This pool is the largest organic pool in natural samples and includes most organic signatures indicative of biological processes. Water-insoluble organics include humic substances, lignin-derived polymers, biofilms, and microbial residues that resist decomposition and can sequester carbon over geological timescales (Schmidt et al., 2011; Keiluweit et al., 2015; Folhas et al., 2025; Stegen et al., 2016). On Mars, mineral-bound macromolecular organic matter, whether of meteoritic, geological, or biogenic origin (Eigenbrode et al, 2018) provides stable targets for life detection, particularly if coupled with molecular and isotopic analyses that may support interpretations of formation and diagenetic processing. In addition, characterizing the insoluble organic inventory could also offer insights into the abiotic or biotic processes responsible for the reworking and recycling of primary organic matter including the mechanisms for the co-evolution of organics with planetary environments and the timescale in which change took place.

**Mission concepts that do not specifically distinguish these pools.** The Icebreaker mission concept proposed drilling into ice-cemented rock and sediments to detect a broad spectrum of organic molecules from bulk samples, including amino acids, peptides, and lipids, using analytical techniques such as bulk sample derivatization-gas chromatography–mass spectrometry (GC-MS) and microarray-based optical sensor for molecular-recognition (McKay et al., 2013; Glass et al., 2024). The Mars Life Explorer (MLE) concept expands upon this approach by targeting mid-latitude ice to analyze solvent-extractable organic compounds with GC-MS, laser desorption-mass spectroscopy, and tunable laser spectroscopy. Neither concepts focus specifically on the detection of aqueous metabolites within Martian subsurface ice. It is unclear that the MLE approach would detect this portion within extracts. Rather, both concepts aim to observe compositions of bulk organics of both soluble and insoluble phases. Future concepts that specifically target aqueous metabolites may reveal evidence of organic carbon availability, potential metabolic substrates, and environmental conditions compatible with life.

**Science Implementation Approaches.** Robust organic detection strategies require integrated instrumentation systems encompassing sample preparation, molecular separation, detection, and spatial analysis techniques (*e.g.,* Brinckerhoff et al., 2022). Sample preparation forms the foundation of organic analysis, enabling liberation, concentration, or chemical modification of target molecules within complex mineral and ice matrices. Physical processing steps such as filtration, grinding, and extraction isolate particulate and dissolved fractions (*e.g.,* Wilhelm et al., 2021), while thermal processing such as pyrolysis decomposes macromolecular organics into volatile compounds amenable to analysis (Eigenbrode et al., 2018). Chemical processing, including buffering, acid or base hydrolysis, (*e.g.,* Radosevich et al., 2019) and derivatization (e.g., Mahaffy et al., 2012; Goesmann et al., 2017; Trainer et al, 2022), increases molecular volatility or detection sensitivity for analytes such as amino acids, organic acids, lipids, and nucleobases (Chou et al., 2021).

Molecular separation techniques enhance specificity and resolution prior to detection. Gas chromatography (GC), a common approach employed for multiple missions to both Mars and Titan, separates volatile and derivatized compounds based on polarity and volatility, while liquid chromatography (LC) separated soluble and more thermally labile molecules has not yet been implemented in flight. Capillary electrophoresis (CE) provides rapid separations of small polar molecules with minimal sample volume requirements and is highly compatible with fluorescence-based detection methods (Mora et al., 2022, Seaton et al., 2023). Mass separation using high-resolution mass spectrometry (HRMS) resolves isobaric interferences and provides exact mass determinations necessary for confident molecular identification in complex organic mixtures (*e.g.,* Arevalo et al, 2018; 2019).

Molecular detection techniques quantify and identify compounds following separation. Mass spectrometry (MS) offers broad-range detection with structural elucidation capabilities via fragmentation patterns of known and unknown compounds (Chou et al, 2021; Arevalo et al., 2019). Laser-induced fluorescence (LIF) enables highly sensitive detection of fluorescently derivatized compounds such as known amino acids, peptides, and nucleotides (Mora et al., 2022, Seaton et al., 2023).

In contrast to individual molecular detections, bulk sample-resolved and spatial-resolved molecular data characterizes different aspects of the organic material encountered. Bulk sample techniques include evolved gas analysis (EGA) (*e.g.,* Eigenbrode et al, 2018), single-point laser desorption mass spectrometry (LDMS), generalize organic carbon detection techniques. Spatially resolved measurements are achieved through array and imaging detection techniques. For example, rasterization-LDMS maps organic distributions at micron scales, revealing chemical heterogeneity and potential cellular structures (Grubisic et al., 2021; Li et al., 2017). Raman spectroscopy provides vibrational signatures that characterize both organic molecules and their mineralogical context, providing insights into preservation environments and molecular associations. (Abbey et al., 2017; Beyssac, 2020). Future instrumentation aims to merge measurement approaches to provide integrated observations to enhance interpretability (e.g. (Mullin et al. 2025).

**The Value of Diverse Triaging Approaches.** Triaging unknown organic chemistry using diverse analytical approaches is essential for robust astrobiological interpretations. Employing a range of sample preparation, molecular separation, detection, and imaging techniques enables the characterization of both labile (*i.e.,* soluble and reactive) and recalcitrant (i.e., insoluble, non-reactive; include macromolecules) organic pools. Such integrated strategies provide molecular-level details to identify specific molecules of possible biological affinity and degradation products, characterize the bulk organic chemistry to assess overall abundance, complexity, and preservation state, and establish spatial and mineralogical context that informs potential processes involved and the degree of organic evolution from abiogenic to prebiotic to fully biotic states (NASEM, 2023; Davila and Eigenbrode, 2024).

Multiple orthogonal analytical modalities strengthen interpretations of signatures that we associate with biology as we know it and don't know it, by constraining possible abiotic sources, validating measurements across distinct techniques, and ensuring flexibility to interrogate diverse organic chemistries encountered in Martian subsurface environments. This holistic triage approach aligns with the priorities articulated in the Mars Exploration Program Analysis Group's SFL-SAG Charter for a medium-class Mars mission focused on extant life detection. By combining science implementation approaches that enable both aqueous and insoluble organic analyses with versatile preparation and detection workflows, future missions will maximize the likelihood of detecting and confidently interpreting potential evidence for life on Mars and with greater organic chemical context to establish confidence levels for the interpretations.

**Acknowledgement:** ChatGPT (OpenAI, San Francisco, CA, July 2025 version) implemented at NASA Goddard Space Flight Center to generate initial drafts of summary paragraphs based on author-provided prompts. All content was critically reviewed, extensively revised, and fact-checked by the lead author before inclusion.